\documentstyle[11pt,newpasp,twoside]{article}
\def\edcomment#1{\iffalse\marginpar{\raggedright\sl#1\/}\else\relax\fi}
\reversemarginpar

\begin{document}
\title{Core-Collapse Supernovae and Neutron Star Kicks}
 \author{Dong Lai}
\affil{Department of Astronomy, Cornell University, Ithaca, NY 14853, USA}

\begin{abstract}
Recent observations have revealed many new puzzles related to 
core-collapse supernovae, including the formation of magnetars and
black holes and their possible GRB connections. We review our current
understanding of the origin of pulsar kicks and supernova asymmetry. It is
argued that neutron star kicks are intimately connected to the other
fundamental parameters of young neutron stars, such as the initial spin and
magnetic field strength.
\end{abstract}

\def\go{\mathrel{\raise.3ex\hbox{$>$}\mkern-14mu
             \lower0.6ex\hbox{$\sim$}}}
\def\lo{\mathrel{\raise.3ex\hbox{$<$}\mkern-14mu
             \lower0.6ex\hbox{$\sim$}}}
\def\simgreat{\mathbin{\lower 3pt\hbox
   {$\rlap{\raise 5pt\hbox{$\char'076$}}\mathchar"7218$}}}
\def\kms{{\rm km\,s}^{-1}}

\section{Introduction}

The subject of supernovae (SNe) has a long history, but the modern era
of SN research really began in 1934 when Baade and Zwicky made the prophetic
suggestion that the death of massive stars, SN explosions and neutron star
(NS) formation are connected events. This suggestion was confirmed
by the discovery of the pulsar in the Crab supernova remnant (SNR) in 1968;
the SN, as is well known, was actually observed in 1054 by Chinese
astronomers.

Today the mechanism of SN explosion remains an unsolved problem. 
Moreover, observations over the last few years suggest that we may actually
know less than we thought about core collapse and explosion of massive stars. 
Here we discuss a small sample of unsolved problems related to
SNe and NS formation, focusing on the problem of NS kicks.

\smallskip
{\it Basic Paradigm for Core-Collapse Supernovae:}
The current paradigm for core-collapse supernovae 
is that they are neutrino-driven (see, e.g., Bethe 1990;
Janka et al.~2001; Burrows \& Thompson 2002 for reviews): 
As the central core of a massive star collapses to nuclear density, it rebounds
and sends off a shock wave, leaving behind a proto-NS. The shock
stalls at several 100's km because of neutrino loss and
nuclear dissociation in the shock. A fraction of 
the neutrinos emitted from the proto-NS get absorbed
by nucleons behind the shock, thus reviving the shock, 
leading to an explosion on the timescale several 100's ms ---  
This is the so-called ``delayed mechanism''. However, 1D simulations with
detailed neutrino transport seem to indicate that neutrino heating of the
stalled shock, by itself, does not lead to a robust explosion
(e.g., Rampp \& Janka 2000; Liebendoerfer et al.~2002). It has been 
argued that neutrino-driven convection in the proto-NS (which 
tends to increase the neutrino flux) and 
that in the shocked mantle (which tends to increase the neutrino heating
efficiency) are central to the explosion mechanism, although it is not
clear how robust of these convections are (see, e.g., 
Mezzacappa et al.~1998; Fryer \& Warren 2002). 
Clearly, in this ``standard model'',
the problem of SN explosion is a quantitative one,
involving 3D radiation (neutrino) (and possibly relativistic) hydrodynamics.

This basic picture of core-collapse SNe, however, is likely to be incomplete. 
There are some obvious puzzles as a result of recent observations:

\smallskip
{\it Formation of Magnetars:} Observations over the last few years 
have revealed two new classes of young NSs, the soft gamma-ray repeaters
(SGRs) and anomalous X-ray pulsars (AXPs). They have many common
characteristics and are mostly likely related to each other (see Kaspi's
contribution in this proceedings). Strong physical arguments suggest that these
are young NSs endowed with superstrong magnetic fields $B\go 10^{14}$~G (see
Thompson 2001). The existence of these magnetars poses some obvious
questions: Under what conditions core-collapse of massive stars will lead to
radio pulsars vs. magnetars? What is the branching ratio? What is the
origin of the NS magnetic field? Does the B-field (in combination with
rotation) play any dynamical role in the explosion? Currently
we do not have firm answers to these questions
(see, e.g., Thompson \& Duncan 1993).

\smallskip
{\it Black Hole Formation and Core-Collapse SNe:}
BHs are also formed in the core collapse of massive stars. 
Recent observations showed that BH formation can 
be accompanied by SN explosion, at least in two cases: The companion of
the BH X-ray binary GRO J1655-40 (Nova Sco) and that of SAX J1819.3-2525 
(V4641 SGR) have high abundance of $\alpha$-elements (Israelian et al.~1999;
Orosz et al.~2001); these $\alpha$-elements can only be produced in 
a SN explosion. Apparently, the companion stars have been polluted 
by material ejected in the SN that accompanied the formation of the
BH primary (see Podsiadlowski et al.~2002). These observations pose a 
host of questions: What are the differences between a SN that made a NS 
and a SN that made a BH? How is the BH formed? We could have a direct 
collapse to BH in which
the shock wave never successfully makes it to induce an explosion, or we
could have a indirect process where a shock wave successfully
makes an explosion and a NS forms temporarily followed by fall-back,
or loss of angular momentum and thermal energy in the proto-NS 
which then collapses to a BH. This indirect process may explain the
the relatively large space velocity of GRO J1655-40.

On a more speculative side, there is possible connection between SN
and the central engine of gamma-ray bursts (GRBs). In the last few
years a growing list of observations suggests that some GRBs are 
connected with the death of massive star and SNe. The famous case
is GRB 980425, which coincided both in time and in position with a 
Type Ic SN1998bw. In at least 3 GRBs (980326,~970228,~000911),
the rebrightening of the optical afterglows about a month after the initial
bursts have been observed and interpreted as the underlying SNe which emerged
when the afterglows faded. Possible emission line features (K$_\alpha$ of Fe,
O, Mg, Si, etc) in several X-ray afterglows at hours-days may also be
an indication of SNe. 
The obvious question is: under what conditions will
the collapse of a massive star
lead to GRB, SN with BH, SN with NS, etc.?

\section{Evidence for Neutron Star Kicks and Supernova Asymmetry}

It has long been recognized that NSs have space velocities much greater 
than their progenitors'.  A natural explanation for such high velocities 
is that SN explosions are asymmetric, and provide kicks to the 
nascent NSs. In recent years evidence for
NS kicks and SN asymmetry has become much stronger, but the origin of
the kicks remains mysterious. The observations that 
support (or even require) NS kicks fall into three categories:

\smallskip\noindent
{\it (1) Large NS Velocities ($\gg$ the progenitors' velocities 
$\sim 30$~km\,s$^{-1}$):} 

$\bullet$ The study of pulsar proper motion give a mean birth 
velocity $200-500$~km\,s$^{-1}$ 
(Lorimer et al.~1997; Hansen \& Phinney 1997; 
Arzoumanian et al~2002), with possibly a significant population having
$V\go 1000$~km\,s$^{-1}$. 

$\bullet$ Observations of bow shock from the Guitar nebula
pulsar (B2224+65) implies $V \simgreat 1000$~km~s$^{-1}$
(Cordes et al.~1993; Chatterjee \& Cordes 2002).

$\bullet$ The studies of NS -- SNR associations have, 
in some cases, implied large NS velocities, up to $\sim 10^3$~km~s$^{-1}$
(e.g., NS in Cas A SNR has $V>330$~km~s$^{-1}$; Thorstensen et al.~2001).

\smallskip\noindent
{\it (2) Characteristics of NS Binaries:} 
While large space velocities can in principle be accounted for
by binary break-up,
many observed characteristics of NS binaries 
can only be explained by intrinsic kicks:

$\bullet$ The detection of geodetic precession in binary pulsar PSR 1913+16 
(Kramer 1998; Wex et al.~1999; see contributions by J. Weisberg and by
M. Kramer).

$\bullet$ The spin-orbit misalignment in PSR J0045-7319/B-star binary,
as manifested by the orbital plane precession 
(Kaspi et al.~1996; Lai et al.~1995) and fast orbital decay (which indicates
retrograde rotation of the B star with respect to the orbit; Lai 1996a;
Kumar \& Quataert 1997).
Similar precession of orbital plane has been observed in PSR J1740-3052
system (see Stair's contribution). 

$\bullet$ High system radial velocity ($430\,\kms$) of X-ray
binary Circinus X-1 
(Tauris et al.~1999). Also, PSR J1141-6545 has $V_{\rm sys}\simeq 125~\kms$
(see contributions by Ord and Bailes).

$\bullet$ High eccentricities of Be/X-ray binaries 
(Verbunt \& van den Heuvel 1995; but see Pfahl et al.~2002).

$\bullet$ Evolutionary studies of NS binary population 
(in particular the double NS systems)
(e.g., Deway \& Cordes 1987; Fryer \& Kalogera 1997; Fryer et al.~1998). 
 
\smallskip\noindent
{\it (3) Observations of SNe and SNRs:}
There are many direct observations of nearby SNe
(e.g., spectropolarimetry: Wang et al.~2000, Leonard et al.~2001;
X-ray and gamma-ray observations and emission line profiles of SN1987A)
and SNRs (e.g., Aschenbach et al.~1995; Hwang et al.~2002) 
which support the notion that SN explosions are not spherically
symmetric.

\section{NS Kick Mechanisms} 

Now we review three different classes of mechanisms for generating 
NS kicks.

\smallskip
\noindent
{\bf Hydrodynamically Driven Kicks} 
 
The first class of kick mechanisms relies on hydrodynamics.
Since the collapsed core and its surrounding mantle are susceptible
to a variety of hydrodynamical (convective) instabilities, one might
expect that the asymmetries in the density, temperature and 
velocity distributions 
associated with the instabilities can lead to asymmetric matter ejection 
and/or asymmetric neutrino emission. Numerical simulations, however, indicate 
that these local, post-collapse instabilities are not adequate to account 
for kick velocities $\simgreat 50$~km~s$^{-1}$.
To produce sufficient kicks, the key is to have global asymmetric 
perturbations in presupernova cores before collapse.

One possible origin for the pre-SN asymmetry is the 
overstable oscillations in the pre-SN core (Goldreich et al.~1996).
The idea is the following. A few hours prior to core collapse, 
the central region of the progenitor star consists of a Fe 
core surrounded by Si-O burning shells and other layers of envelope.
This configuration is overstable to nonspherical oscillation modes.
It is simplest to see this by considering a $l=1$ mode: If we perturb 
the core to the right, the right-hand-side of the shell will be compressed, 
resulting in an increase in
temperature; since the shell nuclear burning rate depends sensitively
on temperature (power-law index $\sim 47$ for Si burning and $\sim 33$ for O
burning), the nuclear burning is greatly enhanced; this generates a large 
local pressure, pushing the core back to the left. 
The result is an oscillating g-mode with increasing amplitude. 
There are also damping mechanisms for these modes, the most important
one being leakage of mode energy: 
Since acoustic waves whose frequencies lie above the acoustic cutoff can
propagate through convective regions, each core g-mode will couple to an
outgoing acoustic wave, which drains energy from the core g-modes.
In another word, the g-mode is not exactly trapped in the core.
Our calculations (based on the $15M_\odot$ and $25M_\odot$
presupernova models of Weaver \& Woosley) indicate that 
a large number of g-modes are overstable, although for low-order modes
(small $l$ and $n$) the results depend sensitively on the detailed
structure and burning rates of the presupernova models (see Lai 2001).
The typical mode periods are $\simgreat 1$~s, the growth time
$\sim 10-50$~s, and the lifetime of the Si shell burning is $\sim$ hours.
Thus there could be a lot of e-foldings for the nonspherical
g-modes to grow. Our preliminary calculations based on the recent models of
A. Heger and S. Woosley (Heger et al.~2001)
give similar results (work in progress).
Our tentative conclusion is that overstable g-modes can potentially grow
to large amplitudes prior to core implosion, although several issues
remain to be understood better.
For example, the O-Si burning shell is highly convective,
with convective speed reaching $1/4$ of the sound speed, and
hydrodynamical simulation may be needed to properly modeled such convection
zones (see Bazan \& Arnett 1998, Asida \& Arnett 2000).

So now we have a way of generating initial asymmetric perturbations
before core collapse. During the collapse, the asymmetries 
are amplified by a factor of 5-10 (Lai \& Goldreich 2000).
How do we get the kick? The numerical simulations by
Burrows \& Hayes (1996) illustrate the effect. Suppose the right-hand-side
of the collapsing core is denser than the left-hand side. As the shock wave 
comes out after bounce, it will see different densities in different
directions, and it will move preferentially on the direction
where the density is lower. So we have an asymmetric shock 
propagation and mass ejection, a ``mass rocket''.
The magnitude of kick velocity is 
proportional to the degree of initial asymmetry in the imploding core.

\smallskip
\noindent
{\bf Neutrino -- Magnetic Field Driven Kicks}

The second class of kick mechanisms rely on asymmetric neutrino emission
induced by strong magnetic fields. 
The fractional asymmetry $\alpha$ in the radiated neutrino energy required to
generate a kick velocity $V_{\rm kick}$ is $\alpha=MV_{\rm kick}c/E_{\rm tot}$
($=0.028$ for $V_{\rm kick}=1000$~km~s$^{-1}$, NS mass
$M=1.4\,M_\odot$ and total neutrino energy radiated $E_{\rm tot}
=3\times 10^{53}$~erg). There are several possible effects:

{\it (1) Parity Violation:} Because weak interaction is parity
violating, the neutrino opacities and emissivities in a magnetized nuclear
medium depend asymmetrically on the directions of neutrino momenta with respect
to the magnetic field, and  this can give rise to asymmetric neutrino emission
from the proto-NS. Calculations indicate that to generate interesting kicks
with this effect requires the proto-NS to have a large-scale, ordered 
magnetic field of at least a few $\times 10^{15}$~G (see Arras \& Lai 1999
and references therein). 

{\it (2) Asymmetric Field Topology:} 
Another effect relies on the asymmetric magnetic field 
distribution in proto-NSs: Since the cross section for
$\nu_e$ ($\bar\nu_e$) absorption on neutrons (protons) depends on the local
magnetic field strength, the local neutrino fluxes emerged from
different regions of the stellar surface are different. Calculations
indicate that to generate a kick velocity of $\sim 300$~km~s$^{-1}$ using this 
effect alone would require that the difference 
in the field strengths at the two opposite poles of the star 
be at least $10^{16}$~G (see Lai \& Qian 1998). Note that 
only the magnitude of the field matters here. 

{\it (3) Dynamical Effect of Magnetic Fields:}
A superstrong magnetic field may also play a dynamical role in the
proto-NS. For example, it has been suggested that a locally strong
magnetic field can induce ``dark spots'' (where the neutrino flux is
lower than average) on the stellar surface by suppressing
neutrino-driven convection (Duncan \& Thompson 1992). While it is
difficult to quantify the kick velocity resulting from an asymmetric
distribution of dark spots, order-of-magnitude estimate indicates that a local
magnetic field of at least $10^{15}$~G is needed for this effect to be
of importance.

\smallskip\noindent
{\bf Electromagnetically Drievn Kicks}
 
Harrison \& Tademaru (1975) show that electromagnetic (EM) radiation 
from an off-centered rotating magnetic dipole imparts
a kick to the pulsar along its spin axis. The kick is attained
on the initial spindown timescale of the pulsar (i.e.,
this really is a gradual acceleration), and
comes at the expense of the spin kinetic energy. A reexamination
of this effect (Lai et al.~2001) showed that the force on the
pulsar due to asymmetric EM radiation is larger than the original 
Harrison \& Tademaru expression by a factor of four.
Nevertheless, to generate interesting kicks using this mechanism
requires the initial spin of the NS to be less than $1-2$~ms.
Gravitational radiation may also affect the net velocity boost.

\section{Astrophysical Constraints on Kick Mechanisms}

The review in \S 3 clearly shows that NS kick is not only a matter of
curiosity, it is intimately connected to the other fundamental
parameters of young NSs (initial spin and magnetic field), and is an 
important ingredient of SN astrophysics. 

One of the reasons that it has been difficult to pin down the kick 
mechanisms is the lack of correlation between NS velocity and the other 
properties of NSs. The situation may have changed with the recent X-ray
observations of the compact X-ray nebulae of 
the Crab and Vela pulsars, which have a two sided asymmetric jet
at a position angle coinciding with the position angle of the pulsar's
proper motion (Pavlov et al.~2000; Helfand et al.~2001). 
The symmetric morphology of the nebula
with respect to the jet direction strongly suggests that the jet is
along the pulsar's spin axis. Analysis of the polarization angle
of Vela's radio emission corroborates this interpretation (Lai et al.~2001).
Thus, while statistical analysis 
of pulsar population neither support nor rule out any 
spin-kick correlation, at least for the Vela and Crab pulsars,
the proper motion and the spin axis appear to be aligned.

The apparent alignment between the spin axis and proper motion
raises an interesting question: Under what 
conditions is the spin-kick alignment expected for different kick 
mechanisms? Let us look at the three classes of mechanisms discussed
before (Lai et al.~2001): (1) For the electromagnetically driven kicks,
the spin-kick slignment is naturally produced. (Again, note that 
$P_i\sim 1-2$~ms is required to generate sufficiently large $V_{\rm kick}$).
(2) For the neutrino--magnetic field driven kicks: The kick is imparted to 
the NS near the neutrinosphere (at 10's of km) on the neutrino diffusion time, 
$\tau_{\rm kick}\sim 10$~seconds. As long as the initial spin period 
$P_i$ is much less than a few seconds, spin-kick alignment is naturally 
expected.
(3) For the hydrodynamically driven kicks: because the kick is imparted at a
large radius ($\go 100$~km), to get effective rotational averaging, we require
that the rotation period at $\sim 100$~km to be shorter than the kick
timescale ($\sim 100$~ms). This translates to $P_{\rm NS}\lo 1$~ms, which means
that rotation must be dynamically important. 

Currently we do not know whether spin-kick alignment is
a generic feature of all pulsars; if it is, then it can provide
powerful constraint on the kick mechanisms and the SN explosion
mechanisms in general.

\smallskip
I thank P. Arras, A. Burrows, D. Chernoff, J. Cordes,
P. Goldreich, A. Heger, and Y.-Z. Qian for collaboration/discussion.
This work is supported by NASA, NSF and the Sloan foundation.

\end{document}